\documentclass{amsart} 
\usepackage{amsmath,amssymb}
\usepackage{graphicx}
\usepackage{url}

\title{Enabling full-speed random access to the entire memory on the A100 GPU}
\author{Alden Walker}

\begin{document}

\begin{abstract} 
We describe some features of the A100 memory architecture.
In particular, we give a technique to reverse-engineer some
hardware layout information.  Using this information, we show
how to avoid TLB issues to obtain full-speed random HBM access
to the entire memory, as long as we constrain any particular thread to a
reduced access window of less than 64GB.
\end{abstract}

\maketitle

\section{Introduction}

\subsection{A100}

The A100 GPU is an accelerator intended for HPC use.
A detailed description of its features as released by NVIDIA
can be found on their technical blog~\cite{blog} and in a whitepaper~\cite{whitepaper}.
There are several versions of the A100; initial GPUs had 40GB of HBM, but the memory
increased to 80GB in some later cards.  For our experiments,
we used a SXM4-80GB GPU.

The A100 is divided into 8 GPCs (graphics processing clusters), each of which has
8 TPCs (texture processing clusters), each of which has 2 SMs (streaming multiprocessors).
In order to increase yield (by accepting slightly defective chips), only 7 GPCs are
made available, each of which has 7 or 8 available TPCs, for a promised total of 108 SMs overall.
The special PTX assembly registers \texttt{\%smid} and \texttt{\%nsmid} allow
the programmer to know which SM a particular thread block is executing on, but
there is no simple way to determine which GPC each SM is on, and this may vary
card to card.

\subsection{TLBs}

A TLB (translation lookaside buffer) is a table which stores the physical locations
of some pages of virtual memory.  If a request for memory hits
one of the pages whose location is recorded in the TLB, then that request can
quickly be served.  If the request lies in a different page, then a more lengthy
process is required.  The amount of physical memory accessible by some compute
may be larger than the amount represented by the entire TLB.  That is, while
that compute may have some amount of ``random access'' memory, it cannot
actually access the entire memory in a random access fashion at full speed.
The ``reach'' of a TLB is the amount of memory represented by the number of
pages it can store.

The technical specifications of the A100 GPU do not mention a TLB.  However,
Section 1.4.3 of the tuning guide~\cite{tuning} remarks on a 64GB NVLink TLB
for incoming remote requests, and it seems that this is not the only 64GB TLB
on the chip (See Section~\ref{sec:basic}).

\subsection{Use case and summary of results}

Suppose we have an application that would like random access to a large portion
of the HBM.  Here ``random access'' still includes the requirement that we
have warp-coalesced accesses of 128 bytes (32 32-bit words).
So ``random cache-line access'' might be more accurate.
Even larger accesses (say, 32 64-bit words) can achieve closer to the
theoretical memory bandwidth, but we don't want to assume that our application can guarantee
such large accesses, and our results are somewhat orthogonal to this concern.

If we have every warp on the device access random cache lines in some region, we find
that performance drops off precipitously if the region is larger than 64GB.
This is unfortunate.  However, we might be able to restructure our application
so that while the total region of accessed memory is larger,
any particular warp only accesses a smaller region.

Doing this naively produces no benefit.  However, by probing the layout
of the GPU, we can determine which SMs share a TLB.  If we
ensure that all the warps in all the SMs in a given group are restricted to a 64GB
window (or, say, half the memory for simplicity), then we can get full-speed
random access to the entire GPU memory.

The TLB probing ideas here are similar to the paper~\cite{karnagel}.

\section{Experiments}

\subsection{Basic benchmark}
\label{sec:basic}

\begin{figure}[ht]
\begin{center}
\includegraphics[width=0.75\textwidth]{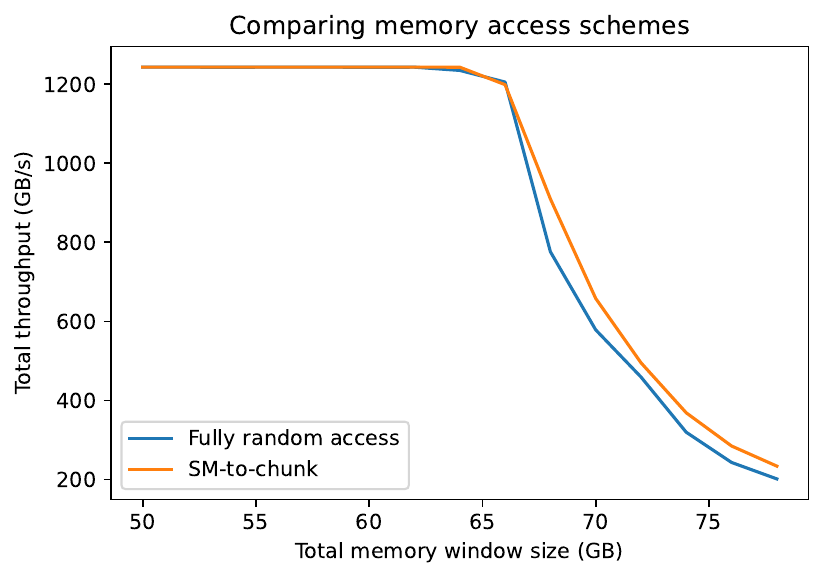}
\end{center}
\caption{Memory throughput for random access.}
\label{fig:baseline}
\end{figure}

To establish a baseline, we first do an experiment in which every warp
reads random coalesced arrays of 32 32-bit words in a memory region of varying size.
We do another experiment in which we divide the memory in half
and for each SM, we pick a random half and require that each warp executing on
that SM reads from the chosen half.  The results of these experiments are shown in
Figure~\ref{fig:baseline}.

Obviously, something dramatic happens when the memory region is larger than
about 64GB.  We remark that the theoretical bandwidth of about 1900 GB/s
is larger than we achieve here.  This is expected, as HBM requires large
transaction sizes to get full efficiency.  Reading 32 64-bit words achieves
about 1400 GB/s, and 32 128-bit words achieves about 1600 GB/s.  We would also
expect sequential reads to be even better.  However, this issue is orthogonal to
the existence of the throughput drop-off at 64GB and our results, so we will not
address it further.

\subsection{Determining SM resource groups}
\label{sec:sm_to_gpc}

\begin{figure}[ht]
\begin{center}
\includegraphics[width=0.75\textwidth]{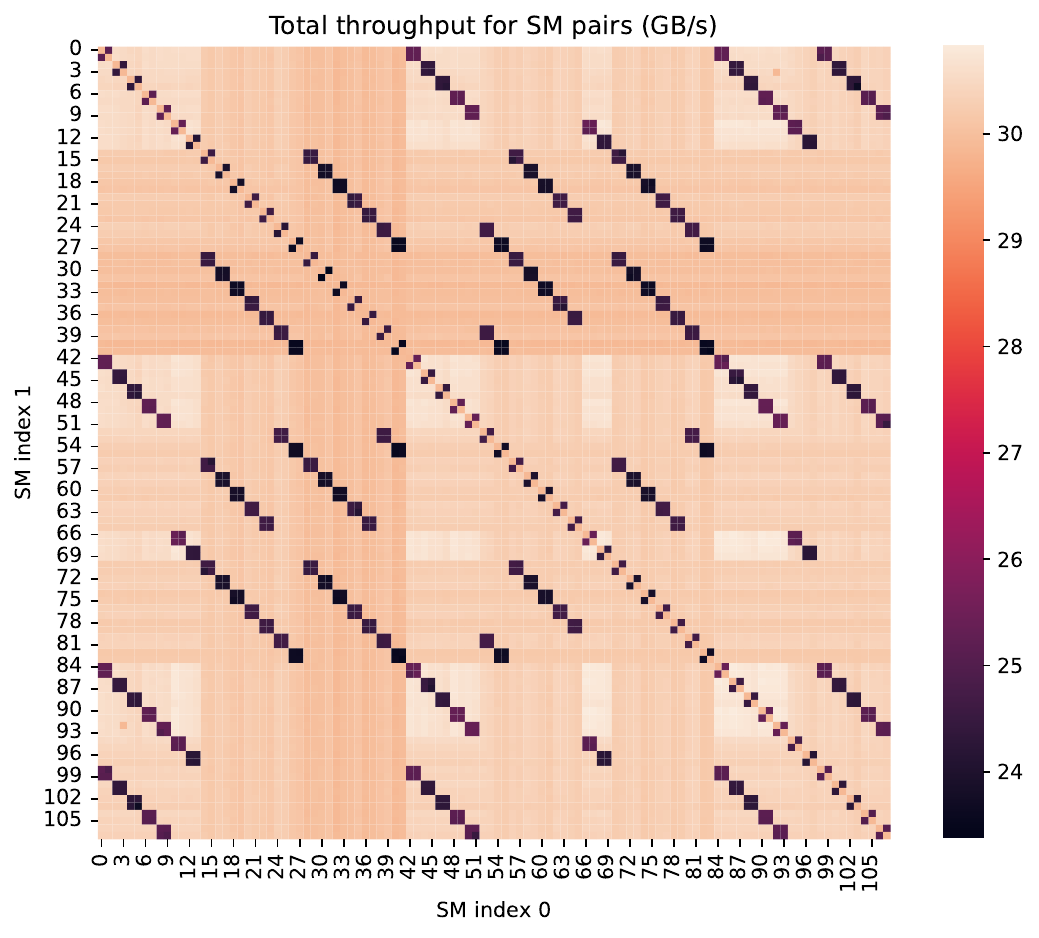}
\end{center}
\caption{Probing SM pairs.}
\label{fig:heatmap}
\end{figure}

The results in Section~\ref{sec:basic} could result from a single
64GB TLB, or they could result from a 64GB TLB per GPC, or something else
entirely.  To get a better understanding, we can do probing experiments
to figure out which SMs share resources.  We expect that if two SMs are
in the same GPC, then they share resources needed to interact with the rest of the
chip (say, for memory or PCIe access).  So if we run our memory access kernel
with only two SMs, we expect that we should see a drop in throughput when
those SMs are in the same GPC as compared to the throughput when they are in
different GPCs.  Indeed, we do see a promising pattern in Figure~\ref{fig:heatmap}

\begin{figure}[ht]
\begin{center}
\includegraphics[width=0.75\textwidth]{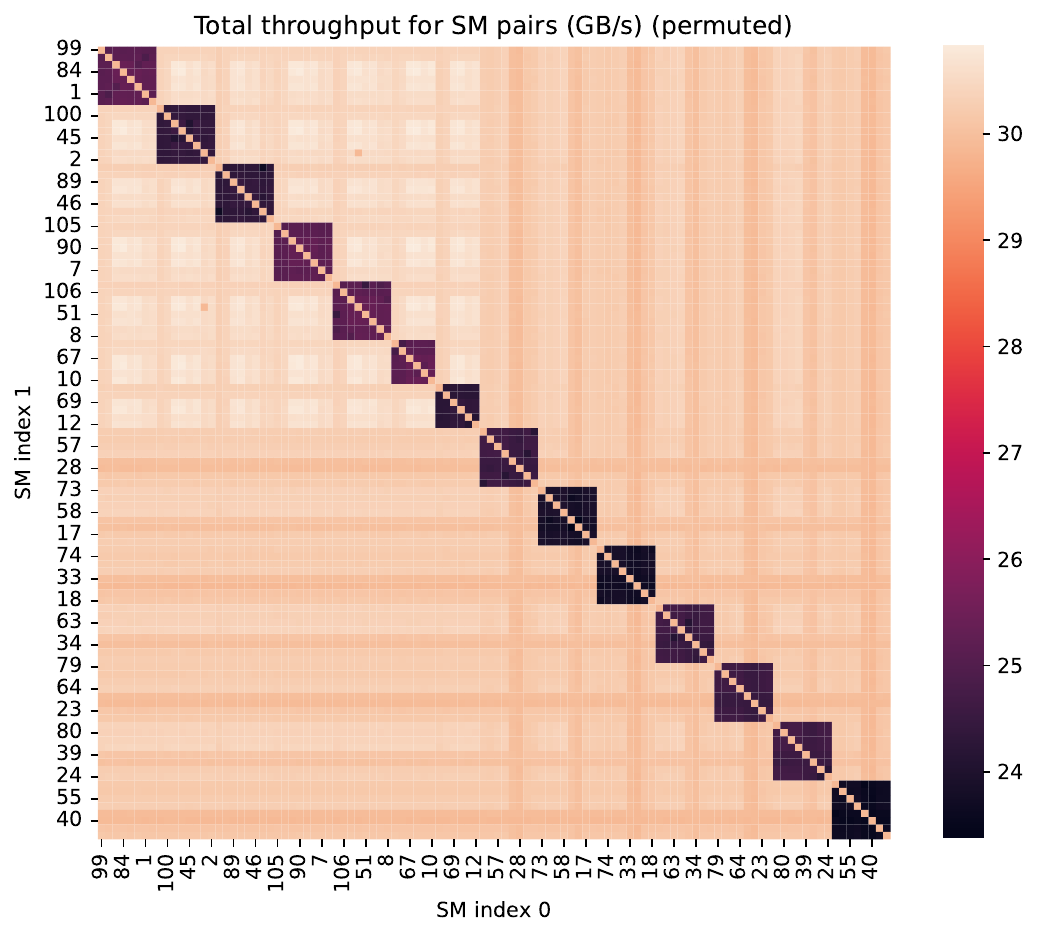}
\end{center}
\caption{Rearranging SM indices to clarify SM memory resource groupings.}
\label{fig:heatmap_sorted}
\end{figure}

Note that the darker boxes in Figure~\ref{fig:heatmap} are actually 2-by-2, the most
likely explanation being that the two SMs within a TPC have consecutive indices.
We are not concerned with the more faint pattern in the background, although it
would be interesting to understand it. By rearranging the indices as
shown in Figure~\ref{fig:heatmap_sorted}, we can more clearly see
how SMs groups share memory access resources.

We know there are 7 GPCs, each with 7 or 8 TPCs.  In Figure~\ref{fig:heatmap_sorted},
there are 14 distinct groups, each with 6 or 8 SMs.  A reasonable explanation
is that each half of each GPC is served by some sort of memory controller.
So in fact the important grouping appears to be finer than simply which GPC 
an SM is in.

\subsection{Checking for resource group independence}
\label{sec:tlb}

\begin{figure}[ht]
\begin{center}
\includegraphics[width=0.75\textwidth]{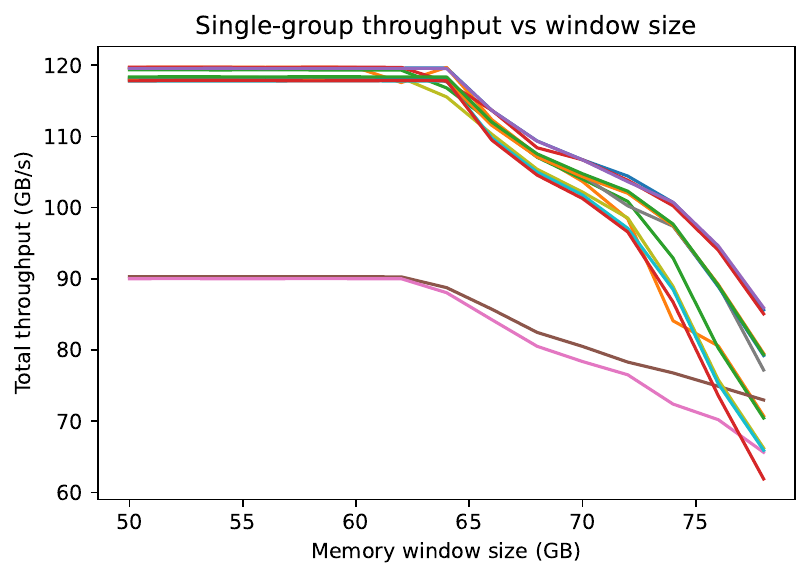}
\end{center}
\caption{Running each resource group individually.}
\label{fig:single_groups}
\end{figure}

If we run each resource group by itself (all warps on all SMs within a given group),
we get Figure~\ref{fig:single_groups}.  The two underperforming groups are the smaller
ones, and indeed the approximate throughput ratio $120/90$ is the ratio of SM counts $8/6$.

\begin{figure}[ht]
\begin{center}
\includegraphics[width=0.75\textwidth]{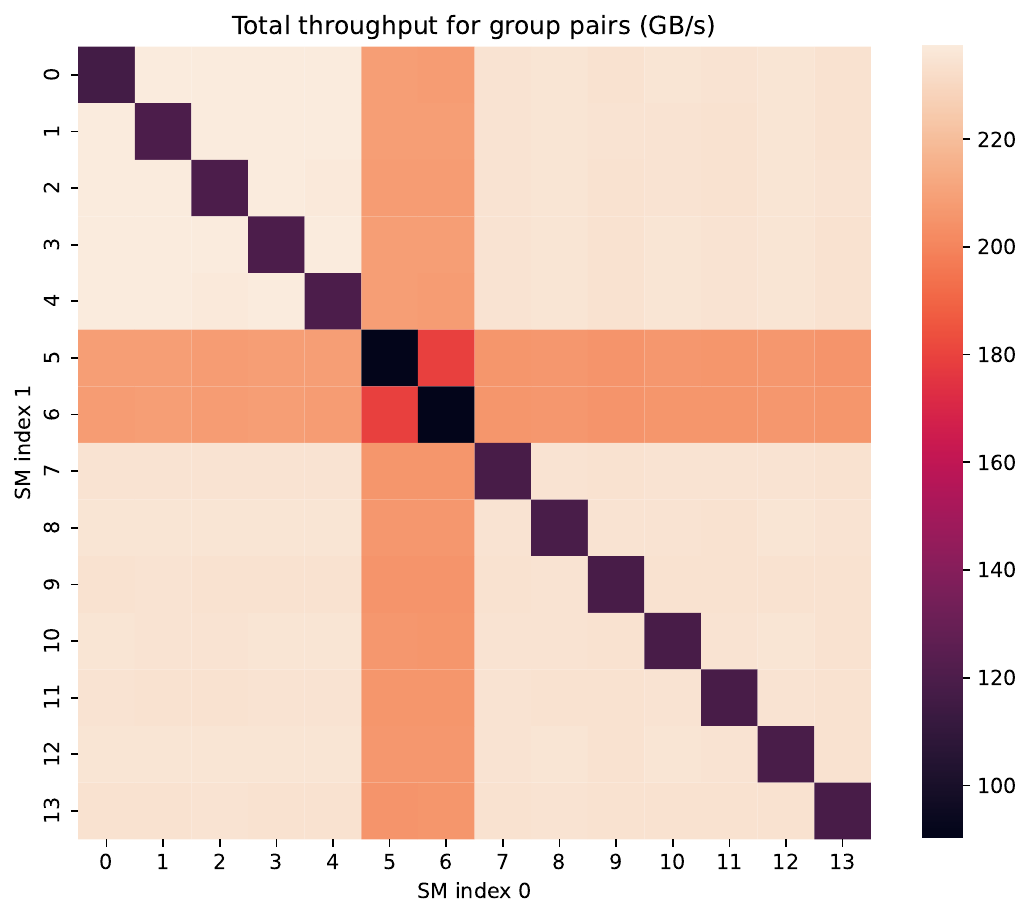}
\end{center}
\caption{Running pairs of resource groups.}
\label{fig:group_pair_heatmap}
\end{figure}

Figure~\ref{fig:single_groups} is not particularly illuminating: it could be
the result of a TLB for each group, or at any higher level.  Although it does
tell us that if there is a higher level TLB, it cannot keep up with even
a small number of SMs. By running two
groups at once, each with random access into a different 40GB region, we
can obtain more information about where the TLB is actually located.
See Figure~\ref{fig:group_pair_heatmap}.

In fact what we see is that the pairs of groups achieve almost exactly double
the throughput of the single groups, indicating that the groups do not seem
to share a TLB.  This suggests that perhaps restricting each group to a region
smaller than 64GB might be enough to avoid all TLBs.

\subsection{Getting full throughput}
\label{sec:full}

\begin{figure}[ht]
\begin{center}
\includegraphics[width=0.75\textwidth]{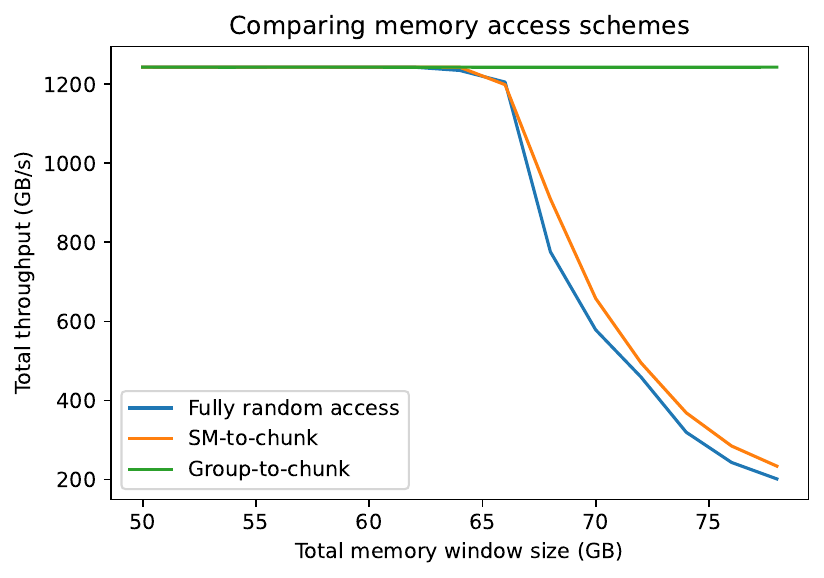}
\end{center}
\caption{Memory throughput for random access, take 2.}
\label{fig:full}
\end{figure}

The natural next step after the experiments in Section~\ref{sec:tlb} is to run
all the groups at once.  That is, to repeat Section~\ref{sec:basic} with an
additional experiment similar to SM-to-chunk which requires all SMs within
a single group to access the same chunk (memory half).
The results are shown in Figure~\ref{fig:full}

Happily, it seems the TLBs are associated with the SM groups, so different
groups can independently access different regions without any penalty for
the total memory region size.

\section{Conclusion}

By probing the SM layout on the A100 GPU, we can determine how the SMs are
served by memory resources.  Apparently, each SM group has its own 64GB TLB.
Thus, by restricting each group to access only within a 64GB window, we can get full-speed
random access to the entire GPU memory.

\end{document}